# Toward a New Science of AI as Cognitive Infrastructure

**Giuseppe Riva** [1-2]

[1] Humane Technology Lab., Catholic University of Sacred Heart, Milan, Italy

[2] Applied Technology for Neuro-Psychology Lab., Istituto Auxologico Italiano, Milan, Italy

**ABSTRACT:**

Contemporary human-AI interaction research currently faces a significant limitation: existing frameworks are inadequate to explain how artificial intelligence systems fundamentally reshape human cognition before conscious awareness occurs. This preprocessing influence, which operates beneath the threshold of deliberate thought, represents a crucial missing layer in the understanding of distributed cognition. This paper introduces "Cognitive Infrastructure Studies" (CIS) as a new interdisciplinary domain to reconceptualize AIs as "cognitive infrastructures". Cognitive infrastructures - e.g., search engines, recommender systems, algorithmic curation platforms, and large language models - exhibit classic infrastructural properties: they are invisible in normal operation, becoming visible only upon breakdown; they are embedded in social and technical arrangements; they are learned as part of membership in digital communities; they link with conventions of practice; and they embody standards that shape what counts as appropriate, relevant, or true. Yet cognitive infrastructures possess distinctive characteristics that distinguish them from traditional infrastructures. Unlike physical infrastructures that passively transport matter or energy, cognitive infrastructures have agency, filtering and curating individuals' perception of reality before it reaches human consciousness. Through narrative scenarios spanning individual (cognitive dependency), collective (democratic deliberation), and societal (governance) scales, we describe how cognitive infrastructures reshape human cognition, public reasoning, and social epistemologies. CIS also provides methodological innovations for studying invisible algorithmic influence: "infrastructure breakdown methodologies", experimental approaches that reveal cognitive dependencies by systematically withdrawing AI preprocessing after periods of habituation.

**Keywords**: Cognitive Infrastructure, Cognitive Infrastructure Studies, epistemic agency, adaptive invisibilty, infrastructure breakdown methodologies

## 1. Introduction

Artificial intelligence is no longer just a set of tools we invoke at will. It now functions as an ambient layer of our information environment, filtering input, directing attention, and framing options invisibly before deliberate reasoning begins. This shift has outpaced existing frameworks. As AI platforms increasingly mediate knowledge, communication, and coordination, we still lack a systematic account of how these hidden, infrastructural processes reconfigure cognition at individual, collective, and societal levels (Riva et al. 2025).

Today's state-of-the-art research only partially captures this phenomenon. For example, researchers found that human–AI feedback loops alter human perceptual, emotional and social judgements (Glickman & Sharot 2025), while communication scholars have documented how social-media algorithms shape public opinion and attention (Milan 2015). Ethicists and sociologists warn of filter bubbles, echo chambers, and the erosion of informed public discourse (Flaxman et al. 2016). As Bommasani et al. (2025) note in their call for evidence-based AI policy, much current research "partially intersects with AI" rather than directly addressing *how AI systems fundamentally reshape human cognitive processes*. However, these perspectives typically frame AI systems as external tools or platforms, deliberately used to enhance cognition. They fail to conceptualize AI as an integrated component of the cognitive environment, operating at a level beneath awareness.

To address this gap, this paper reintroduces and develops the concept of System 0, I recently described with my colleagues in "Nature Human Behaviour" (Chiriatti et al. 2024). Unlike Kahneman's (2011) well-known dichotomy of System 1 (fast, intuitive thinking) and System 2 (slow, analytical thinking), System 0 refers to an invisible, non-human layer of distributed cognition that fundamentally precedes both Systems. It resides not within the individual brain, but in the AI-mediated infrastructure that frames experience before conscious processing begins. System 0 encompasses the algorithmic processes of search engines, recommendation systems, content-ranking platforms, and language models, systems that autonomously sort, prioritize, and pre-select information long before it reaches conscious awareness. These AI infrastructures interact dynamically with user behavior and gradually become functionally integrated into decision-making, attention, memory, and social reasoning (they are more persuasive than humans: Salvi et al. (2025))

What makes System 0 truly novel is its location outside the individual mind (Chiriatti et al. 2025), embedded in socio-technical environments (Baxter & Sommerville 2010) that operate beneath the threshold of reflection. The concept builds on established theories of distributed cognition (Hollan et al. 2000) and the "extended mind thesis" (Clark & Chalmers 1998). Just as we don't hesitate to call a calculator part of our cognitive process when solving complex math problems, AI systems that autonomously filter, prioritize, and pre-process information before it reaches consciousness become functionally integrated into human cognitive processes. The infrastructure is "cognitive" because it performs cognitive functions - reasoning, pattern recognition, relevance judgment - even though it operates outside the biological brain. Moreover, unlike traditional tools that we pick up and put down (e.g., a calculator), a cognitive infrastructure is continuously active, invisibly shaping every information encounter.

System 0 embodies a fundamental tension. Analytically, it functions as an interactional process: a continuous feedback loop between algorithmic filtering and human behaviour that shapes experience before awareness. Simultaneously, AI is evolving into cognitive infrastructure where individuals "think after the machine." What begins as scaffolding can slide into substitution, positioning System 0 as precursor to human Systems 1 and 2. Our project views System 0 as an experimental construct that can be operationalised and measured today.

## 2. The emergence of a new Framework: Cognitive Infrastructure Study - CIS

This challenge demands a new scientific platform (Li Vigni 2021), a meeting point between different specialties pursuing together shared socio-epistemic objectives; *Cognitive Infrastructure Studies (CIS)*. CIS emerges from the recognition that existing disciplinary boundaries cannot adequately capture the complex interplay between AI systems, human cognition, and social coordination that System 0 represents. It introduces a fundamental paradigm shift: from examining individual interactions with AI systems to interrogating the infrastructural conditions under which cognition itself is shaped, distributed, and governed. In this section we will see that CIS emerges from the theoretical convergence of five distinct academic traditions (see Table 1).

| Theoretical Frameworks | Core Contribution to System 0 and CIS | How System 0 Extends Beyond State-of-the-Art |
|---|---|---|
| **Science & Technology Studies (STS)** | Establishes **socio-technical integration and agency distribution** across human-AI networks | **System 0 uniquely focuses on cognitive coupling** rather than general socio-technical relations, specifically examining how AI becomes embedded in reasoning processes below conscious awareness |
| **Cognitive Science** | Provides **theoretical foundation for extended, distributed, and coupled cognition** across human-AI partnerships | **System 0 applies cognitive extension specifically to AI infrastructure**, creating "collective cognitive ecology" where algorithmic preprocessing becomes constitutive of baseline reasoning systems |
| **Digital Sociology** | Reveals **platform-mediated social transformation** and algorithmic life structures | **System 0 conceptualizes platforms as cognitive infrastructure** rather than social infrastructure, focusing on how algorithmic mediation shapes epistemic processes and reasoning quality |
| **Infrastructure Studies** | Explains **invisible operation, classificatory functions, and transparency dynamics** | **System 0 identifies AI as cognitive infrastructure** specifically, emphasizing how algorithmic preprocessing operates as fundamental utility for reasoning, not just information management |
| **Argumentative Theory** | Demonstrates **reasoning's fundamentally social character** and epistemic foundations | **System 0 reveals how AI fragments shared epistemic foundations**, disrupting the social basis of reasoning by creating personalized reality environments that undermine collective deliberation |

**Table 1**. The five distinct academic traditions that provide the theoretical frameworks for CIS

This synthesis establishes both System 0 and Cognitive Infrastructure Studies as rigorously grounded in established scholarship while extending beyond current state-of-the-art to address the pressing contemporary challenge of human-AI cognitive coupling.

### 2.1 Science and Technology Studies (STS): The Socio-Technical Nexus

Cognitive Infrastructure Studies (CIS) explicitly builds on traditions in Science and Technology Studies (STS), grounded in the idea that technological systems and human social systems form an integrated whole. Baxter and Sommerville (2010) introduced the foundational principle that technological systems and human social systems form an integrated whole. System 0 exemplifies this fusion, representing the seamless integration of human cognitive processes with algorithmic operations.

Actor-Network Theory, particularly Latour's (2005) *Reassembling the Social* and Callon's (1986) translation processes, provides crucial theoretical support for understanding how System 0 distributes agency across heterogeneous networks. The principle of generalized symmetry between human and non-human actors establishes that AI systems are not merely tools but active participants in cognitive networks. Law and Hassard's (1999) post-ANT developments further demonstrate how these networks stabilize through ongoing translation processes that render certain connections invisible and automatic.

Practice theory, drawing from Reckwitz (2002) and Schatzki (2002), establishes how System 0 operates through the embedding of AI-mediated processes within routine social practices. Technologies become constitutive of practices through repeated performance that integrates bodily activities, mental activities, materials, and background knowledge. This theoretical foundation explains how System 0 functions below conscious awareness through the gradual habituation of AI-mediated cognitive routines that become seamlessly integrated into everyday thinking practices.

CIS reconceptualizes AI systems not as discrete tools but as "cognitive infrastructures" (Antikythera 2025) that function as foundational, often invisible socio-technical systems shaping how knowledge is created, validated, and circulated. This perspective acknowledges that AI's impact transcends purely technological effects, being deeply linked with human s, organizational processes, and societal norms.

### 2.2. Cognitive Science: Extending the Mind and Mediating Activity

Cognitive science provides the core theoretical architecture for System 0 through established frameworks of extended, distributed, and coupled cognition. Clark and Chalmers' (1998) extended mind thesis establishes the fundamental principle that external processes can become constitutive parts of cognitive systems when they are constantly accessible, automatically endorsed, and easily retrievable. System 0 builds directly on this framework by recognizing all the external cognitive supports (technologies, media, AI systems) as integral components of our baseline reasoning system. When individuals interact with AI systems like ChatGPT or search engines to process information, these interactions exemplify System 0 in operation: an expanded cognitive system comprising person-plus-technology partnerships.

Hutchins' (1995) foundational work on distributed cognition demonstrates how cognitive processes propagate across representational media within sociotechnical systems, with cognitive properties depending on the physical properties of these representational media. The framework of representational state propagation provides the theoretical foundation for understanding how System 0 coordinates cognitive processes across human-AI networks. Hollan, Hutchins, and Kirsh's (2000) extension to human-computer interaction establishes the theoretical precedent for AI-mediated distributed cognition.

Cole and Engeström's (1993) activity theory perspective contributes crucial insights into how System 0 operates through cultural mediation of cognitive activity. Their framework demonstrates how cognitive processes are fundamentally mediated by cultural tools and practices, providing theoretical support for understanding AI systems as sophisticated cultural mediating tools that shape and are shaped by human activity. CIS incorporates this idea by treating the network of people, information resources, and technological artifacts around an individual as part of the fundamental cognitive system in which reasoning takes place. This creates what we term the collective cognitive ecology: the shared ecosystem of minds, media, and machines within which individual cognitive processes are embedded and distributed.

Contemporary cognitive science research establishes human-AI cognitive coupling as a legitimate research domain. Risko and Gilbert's (2016) cognitive offloading framework provides theoretical foundation for understanding how System 0 optimizes mental resource allocation through strategic externalization of cognitive demands. Their research demonstrates that cognitive offloading, "the use of physical action to alter the information processing requirements of a task so as to reduce cognitive demand", is guided by metacognitive evaluations that can be potentially erroneous, leading to suboptimal offloading behavior. This framework illuminates how System 0 functions through dynamic negotiations between internal cognitive resources and external AI capabilities.

### 2.3 Digital Sociology: Algorithmic Mediation and Platform Society

Digital sociology provides crucial theoretical frameworks for understanding how System 0 operates within contemporary platform-mediated social life, particularly through the notion of the platform society (van Dijck, Poell, & de Waal, 2018). The platform society describes how digital platforms have fundamentally penetrated the heart of societies, becoming integral socio-technical infrastructures that actively shape social interactions, learning processes, and decision-making practices rather than serving as passive tools.

Van Dijck et al. (2018) conceptualize platforms as complex socio-technical systems characterized by "complex interactions between users, practices, technologies, and business models, a combination of human and nonhuman actors." This theoretical foundation emphasizes that platforms function as programmable architectures designed to organize interactions among heterogeneous users while systematically collecting, processing, and monetizing data flows. Rather than neutral intermediaries, platforms actively mediate and organize social life through algorithmic curation that determines information visibility, social connections, and knowledge circulation. In the System 0 framework, these pervasive digital structures constitute the foundational cognitive infrastructure that supports and influences human reasoning processes at a societal scale, creating what scholars term a "connective world" where human cognition and algorithmic processing become increasingly intertwined.

Bucher's (2018) concept of "algorithmic life" provides theoretical foundation for understanding how System 0 mediates daily existence through AI systems that shape information flows, social connections, and cognitive practices. Her analysis demonstrates how algorithms function as "powerful brokers of information, communication and society" that become integrated into social practices, aligning precisely with System 0's operation as a below-awareness mediating layer that filters, ranks, and organizes cognitive inputs.

Couldry and Hepp's (2018) "deep mediatization" framework establishes theoretical precedent for understanding how digital media fundamentally reconstruct social reality. Their materialist approach provides methodological foundation for studying System 0's typically invisible operations through empirical investigation of mediated practices that reveal how algorithmic systems become embedded in everyday cognitive routines. This framework shows how mediatization extends beyond mere media use to encompass fundamental changes in how social reality is constructed and experienced.

In sum, the digital sociology perspective emphasizes that powerful socio-technical actors—platforms and their algorithmic systems—now fundamentally structure how knowledge is produced, circulated, validated, and consumed throughout society. CIS leverages this insight by conceptualizing these algorithmic interventions not as external tools but as constitutive elements of contemporary cognitive systems themselves. This represents a transformation toward what we might call "platform-mediated cognition," where human reasoning processes become systematically coupled with algorithmic processing capabilities, creating hybrid cognitive ecologies that operate across individual, social, and technological scales of analysis.

### 2.4 Infrastructure Theory: Invisible Classification and Transparency

Infrastructure studies provide essential theoretical frameworks for understanding System 0's invisible operation and classificatory functions. Bowker and Star's (1999) *"Sorting Things Out"* establishes classification systems as "the scaffolding of information infrastructures" that shape cognition while remaining largely invisible. Their central insight demonstrates that classification systems and infrastructures do not merely reflect existing social organization but actively reshape people's lives, behaviors, and possibilities for action, typically operating below the threshold of conscious awareness. Classification systems function as powerful technologies that can be rendered invisible and naturalized within social and political infrastructures, influencing how information is categorized, prioritized, and accessed in ways that "torque" cognitive processes. Infrastructure tends to become "naturalized" and transparent until breakdown moments reveal its constructed character (Star 1999), meaning we internalize infrastructural influence as part of everyday cognitive background. This principle of "infrastructural inversion" provides methodological approaches for studying System 0's typically invisible functions through systematic ethnographic investigation of breakdown moments when algorithmic mediation becomes temporarily visible. Within the System 0 concept, these ideas imply that the underlying technical systems and standards (from internet protocols to data classifications and AI algorithms) form an unseen layer of cognitive support and constraint.

Star and Ruhleder's (1996) eight characteristics of infrastructure provide comprehensive theoretical framework for analyzing System 0. These characteristics include embeddedness within other structures and social arrangements; transparency during normal operation; reach and scope across multiple sites and temporal scales; acquisition through membership in communities of practice; linkage with conventions of established practices; embodiment of standards and classificatory principles; dependence on installed base of existing systems; and visibility upon breakdown when normal functioning is disrupted.

Recent work by Plantin et al. (2018) demonstrates how digital platforms evolve from discrete applications into foundational infrastructures that support entire ecosystems of social and economic activity. Ananny and Crawford's (2018) analysis of algorithmic transparency reveals why System 0 necessarily resists conventional approaches to algorithmic accountability, highlighting the fundamental tensions between infrastructure's functional invisibility and demands for democratic oversight.

In sum, recognizing System 0 emphasizes that AI-driven cognitive preprocessing has become a fundamental utility of contemporary life, as essential and typically unnoticed as electrical grids or water systems, yet with direct and profound consequences for how societies think, learn, and make collective decisions. This infrastructural perspective reveals System 0 not as an external tool but as embedded socio-technical architecture that constitutes the foundational conditions within which contemporary cognition operates.

### 2.5 Argumentative Theory of Reasoning (Social Cognition)

The final theoretical component of CIS derives from understanding human reasoning as inherently social. Traditional views conceptualized reasoning (Kahneman's System 2) as solitary logical faculty, but Mercier and Sperber's argumentative theory (Mercier & Sperber 2011, 2017) reframes it as fundamentally social interaction tool. According to their influential 2011 work, reasoning's primary function is argumentative, "to devise and evaluate arguments intended to persuade", rather than pursuing abstract truth in isolation. Humans evolved to reason *with and for others* because survival depends on communication and cooperation. This theory explains why people exhibit confirmation bias and reasoning fallacies when alone, yet perform better in group discussions where they critique each other's arguments. As Mercier and Sperber (2017) emphasize in *The Enigma of Reason*, reasoning functions as "first and foremost a social competence" for producing and evaluating justifications in communal settings. We reason to explain ourselves and convince others, while scrutinizing others' claims for acceptance. System 0's theoretical framework leverages this insight: effective reasoning quality depends fundamentally on social interaction and shared epistemic foundations. While people may disagree about interpretations, productive democratic deliberation requires citizens operate from common basic facts about reality.

System 0 fundamentally disrupts this prerequisite by fragmenting shared reality into personalized information environments. Instead of encountering identical basic facts through common sources, AI algorithms now curate different reality versions for different users, optimized for engagement rather than accuracy or democratic discourse. This transformation elevates System 0 from an individual concern to a systemic threat to democratic deliberation and societal cohesion, as AI preprocessing may generate a new "Digital We" (Riva 2025a) disrupting the open-endedness of human culture (Morgan & Feldman 2024).

## 3. The Infrastructural Turn in Cognition

The convergence of Science and Technology Studies, cognitive science, digital sociology, infrastructure theory, and argumentative theory provides unprecedented theoretical depth for understanding how AI systems become embedded in the basic fabric of human thought and social interaction. While each theoretical tradition contributes essential conceptual foundations, Cognitive Infrastructure Studies transcends these individual contributions by synthesizing them into a comprehensive theoretical framework.

Specifically, CIS reconceptualizes AI not as discrete tools, but as "cognitive infrastructures". These are foundational, often invisible systems that process, filter, and transform information before it reaches human consciousness. Drawing from infrastructure studies (Bowker & Star, 1999; Larkin, 2013), cognitive infrastructures exhibit classic infrastructural properties: they are invisible in normal operation, becoming visible only upon breakdown; they are embedded in social and technical arrangements; they are learned as part of membership in digital communities; they link with conventions of practice; and they embody standards that shape what counts as appropriate, relevant, or true (Susan Leigh Star & Karen Ruhleder 1996).

Yet cognitive infrastructures possess distinctive characteristics that distinguish them from traditional infrastructures. Unlike physical infrastructures that transport matter or energy, cognitive infrastructures have agency (Floridi 2025) filtering and curating individuals' perception of reality before it reaches human consciousness.

Cognitive infrastructures possess distinctive characteristics that differentiate them from traditional physical infrastructures:

- **Anticipatory Personalization:** They operate through what can be termed "anticipatory personalization," constantly learning from behavioral patterns to predict and shape future cognitive states (Chen et al. 2021). This proactive shaping means AI is not merely reacting to user input but actively guiding the user's cognitive trajectory.
- **Adaptive Invisibility:** A critical feature is their "adaptive invisibility". The more sophisticated and integrated these systems become, the more seamlessly they blend into cognitive workflows, making their pervasive influence increasingly difficult to detect or resist (Susser 2019). This characteristic poses a significant challenge for empirical study, as users often remain unaware of the extent to which their cognitive processes are being mediated.
- **Automation of Relevance Judgment:** One of the most significant features of cognitive infrastructure is the automation of "relevance judgment" (Coeckelbergh 2025). This cognitive task, which once belonged solely to human attention and will, is now frequently delegated to machines (Mittelstadt et al. 2016).
- **Shift in Locus of Epistemic Agency:** Consequently, this automation leads to a profound shift in the "locus of epistemic agency" (Nguyen 2020; Nieminen & Ketonen 2024). Decisions about what is worth knowing, seeing, or acting upon are increasingly performed by non-human systems, fundamentally altering the landscape of knowledge acquisition and decision-making.

A predictable objection argues that these characteristics are not unique to cognitive infrastructures, pointing out that both digital and pre-digital technologies, from search algorithms and social media platforms to libraries, newspapers, and broadcast media, have similarly mediated knowledge access, shaped attention, and influenced what information reaches users. However, this objection fundamentally misunderstands the revolutionary nature of cognitive infrastructure. The necessity of this concept lies not in the novelty of individual features, but in their unprecedented convergence, scale, and systemic integration within AI-mediated environments (see Table 2 below).

| Dimension | Pre-AI / Traditional Systems | AI Cognitive Infrastructures | Why It Matters (Novelty) |
|---|---|---|---|
| **Qualitative Transformation** | Static filters, explicit queries, slow/manual adaptation | Massive-scale algorithmic curation, real-time, self-modifying | Represents a qualitative leap requiring new analytical and governance tools |
| **Integration** | Isolated tools used intentionally and episodically | Continuous, ambient cognitive layer embedded across daily environments | Creates a systemic, invisible environment shaping what can be known, debated, or acted upon |
| **Agency** | Human-controlled anticipation and relevance | Delegation of priority-setting and attention allocation to non-human agents | First large-scale transfer of epistemic agency to autonomous systems |
| **Emergent Properties** | Independent, domain-specific effects | Reinforcing loops of dependency, conditioning, and epistemic stratification | Produces system-level effects reshaping cognition itself |
| **Cross-Domain Reach** | Separate silos (search, news, commerce, social) | Interconnected ecosystems sharing data across domains | Generates unified environments and network effects that amplify influence and centralize control |

**Table 2**. The main characteristics of AI as Cognitive Infrastructure

This infrastructural role becomes critical when viewed through Mercier and Sperber's Argumentative Theory (Mercier & Sperber 2011, 2017), which, as we just hve seen, ademonstrates that human reasoning evolved primarily "to devise and evaluate arguments intended to persuade" rather than pursue objective truth in isolation. Their research reveals reasoning is fundamentally social and argumentative: it works best when people with different perspectives argue from shared factual foundations. The crucial insight is that effective collective reasoning depends on shared epistemic foundations. While people may disagree about interpretations, productive democratic deliberation requires citizens operate from common basic facts about reality.

System 0 fundamentally disrupts this prerequisite by fragmenting shared reality into personalized information environments. Instead of everyone encountering the same basic facts through common sources, AI algorithms now curate different versions of reality for different users, optimized for engagement rather than accuracy or democratic discourse. The synergistic interplay between "anticipatory personalization" and "adaptive invisibility" creates a powerful, self-reinforcing loop. AI systems learn user preferences and behaviors (Binz et al. 2025) and subtly steer them (Kramer et al. 2014; Werner et al. 2024), while simultaneously becoming less perceptible to the user.

This transformation elevates System 0 from an individual concern to a systemic threat to democratic deliberation and societal cohesion, as AI preprocessing may generate a new "Digital We" (Riva 2025a) disrupting the open-endedness of human culture (Morgan & Feldman 2024).

For these reasons, the category of cognitive infrastructure is essential. It captures a system-level transformation in which AI does not merely assist cognition but constitutes the infrastructural conditions of thought itself. Without this category, we risk underestimating both their epistemic power and the governance challenges they present (Lorenz-Spreen et al. 2023; NMI 2025; Salvi et al. 2025).

CIS highlights how algorithmic systems stratify knowledge access, exacerbating existing inequalities (Benkler 2006; van Dijck et al. 2018) in epistemic resources and civic participation. Just as societies invest in and regulate physical infrastructures for transportation, sanitation, and energy, because they are vital to public life, so too must we interrogate and govern cognitive infrastructures because they:

- **Structure fundamental cognitive capacities** (not just transit but thought);
- **Shape information environments across populations**, not just individuals;
- **Generate behavioral dependencies**, making disengagement impractical;
- **Amplify network effects**, centralizing cognitive influence;
- **Require regulatory oversight**, especially when privatized;
- **Demand shared standards**, ensuring transparency, accessibility, and interoperability.

By foregrounding these dynamics, CIS invites a shift in digital governance: from regulating *individual AI applications* to managing the underlying cognitive architectures that increasingly define how societies know, decide, and evolve. It calls for new analytic tools, empirical research, and policy frameworks that account for System 0's pervasive, infrastructural role in mediating cognition and power.

In sum, Cognitive Infrastructure Studies proposes a paradigm for understanding AI's societal role that is both conceptually ambitious and deeply grounded in social theory. It does not discard existing models of human-AI interaction but instead synthesizes and extends them to address the infrastructural and cognitive transformations defining the digital age

## 4. Living with System 0: How Cognitive Infrastructures Reshape Human and Societal Thinking

Understanding the reach of cognitive infrastructure requires tracing its effects across different scales, from the individual mind to collective social dynamics to the societal systems that govern them. Through narrative scenarios, we can better grasp the transformative, and at times troubling, implications of life under the guidance of System 0.

### 4.1 The Individual Scale: Dependency, Identity, and the Reshaping of Thought

Sarah paused mid-sentence during her presentation, troubled by the realization that it had been a long time since she had completed a complex thought unaided. Her morning had started—as most days did—with an AI-curated digest shaping her interpretation of world events. Her queries to the web were completed before she fully formed them. Her calendar, emails, and documents were pre-tagged, pre-sorted, and pre-analyzed.

What initially felt like convenience had quietly become cognitive infrastructure. Over time, Sarah's thinking rhythms had synchronized with the cadence of algorithmic assistance. Her attention span adjusted to the tempo of feeds and alerts, her reasoning outsourced to suggestions offered just-in-time. Without noticing, her internal search for meaning had become externally scaffolded.

More than efficiency was at stake. These systems were shaping *who* she thought she was. The articles she read, the videos she watched, even the topics she considered important, were increasingly channeled through invisible filters. Her preferences reflected back to her in a self-reinforcing loop of algorithmic affirmation, narrowing her intellectual world. Her learning style shifted too: from patient knowledge construction to fluent prompt engineering and hyper-efficient synthesis. She gained new skills, but wondered what she had lost.

Sarah's story illustrates a profound shift in human cognition: AI systems are not just tools we use; they are becoming environments we inhabit. This raises questions not only about dependency but about identity, agency, and the evolving shape of the human mind in an AI-saturated world.

### 4.2 The Collective Scale: Democracy in the Age of Algorithmic Mediation

Mayor Chen had high hopes for her town's new AI-powered civic engagement platform. It promised to deepen participation, surface diverse voices, and support evidence-based discussion. And on the surface, it delivered—more residents were joining town forums than ever before.

Yet, as the first live session unfolded, Chen watched democracy take an unexpected turn. The system's engagement algorithms favored posts that elicited strong reactions over those that offered complex ideas. Nuanced policy proposals sank in favor of emotional soundbites. Micro-communities formed quickly, but around narrow interests, fragmenting the public sphere into algorithmically curated echo chambers.

The same infrastructures that enabled participation were subtly shifting the terms of engagement. The forum no longer resembled a deliberative agora, but a reactive marketplace of attention. Worse, the AI's content curation began shaping the town's *memory*, certain local traditions and issues were elevated or erased based not on their civic value but their engagement potential.

Chen's experience reveals how cognitive infrastructures shape not just individual thought, but collective reasoning. What communities consider "urgent," "true," or "representative" is increasingly mediated by systems optimized for engagement rather than epistemic richness or democratic integrity. These systems don't merely host civic life, they configure its conditions.

### 4.3 The Societal Scale: Governance, Equity, and Cognitive Justice

Commissioner Martinez scrolled through the cognitive infrastructure dashboard with growing alarm. The national audit revealed stark disparities in algorithmic access and literacy. In wealthier districts, AI-enhanced tools helped students personalize learning and professionals optimize decision-making. In less connected regions, outdated systems and minimal training left communities behind, not just economically, but cognitively.

A new form of inequality was emerging: *cognitive inequity*. It wasn't just about who had devices or internet, it was about who had access to infrastructures that structured thinking itself. This raised a daunting question: How do you regulate the very conditions under which people form beliefs, evaluate evidence, or make sense of the world?

Martinez and her team confronted an unprecedented governance dilemma. The leading platforms resisted scrutiny, citing proprietary algorithms and trade secrets. Yet these systems held sway over population-level cognition. Calls for transparency were met with claims of technical incomprehensibility. At stake was nothing less than human agency itself, early research showed that AI-saturated environments could subtly reduce independent reasoning and increase behavioral conformity.

To respond, Martinez proposed a new paradigm: treat cognitive infrastructure like public utilities, essential services that must be designed, maintained, and regulated in the public interest. Governance models would need to prioritize cognitive equity, transparency, and human flourishing over mere efficiency. Otherwise, society risked creating a future where the architectures of thought served profit more than people.

### 4.4 Beyond Awareness, Toward Design

Together, these scenarios dramatize how AI-driven preprocessing, System 0, reshapes human cognition, collective reasoning, and societal functioning in invisible but foundational ways. Sarah's cognitive dependencies, Chen's fragmented public sphere, and Martinez's policy dilemmas are not isolated anecdotes but signals of a broader transformation.

To understand, and ultimately govern, this transformation, researchers must develop tools and methods to surface the unseen, to make visible the architectures of cognition that now structure our lives. The challenge is not just to observe these systems, but to imagine how they might be reconfigured to support cognitive diversity, epistemic justice, and democratic integrity in a world increasingly shaped by invisible minds.

## 5. Methodological Imperatives for CIS

The empirical study of cognitive infrastructures presents significant methodological challenges, particularly in capturing the subtle, preconscious effects of algorithmic preprocessing across individual, collective, and cultural scales. Traditional social science methods, such as surveys, interviews, and ethnographies, remain indispensable for eliciting conscious beliefs, attitudes, and meaning-making practices. However, these techniques are fundamentally limited in their ability to detect invisible algorithmic mediation, precisely because they rely on participants' explicit awareness and reflective accounts. Likewise, most laboratory-based human–AI interaction studies tend to focus on conscious tool use or deliberate collaboration, overlooking the ambient, pre-reflective integration of AI systems into everyday cognition.

To address this epistemic gap, Cognitive Infrastructure Studies (CIS) advances what can be defined "infrastructure breakdown methodologies": experimental designs that make the invisible cognitive role of AI systems visible by simulating conditions of algorithmic withdrawal, degradation, or disruption. These methodologies are inspired by the core insight from infrastructure studies—particularly Star and Bowker's foundational claim that "infrastructure becomes visible upon breakdown" and transpose this logic from material systems (roads, cables, databases) to cognitive environments. In practice, these methods involve embedding participants in seemingly "neutral" digital contexts where AI preprocessing (e.g. algorithmic filtering, automated summarization, recommendation ranking) operates invisibly in the background, shaping information flow, relevance hierarchies, and decision scaffolding. After a period of habituation, algorithmic supports are systematically altered or removed, allowing researchers to observe patterns of performance degradation, strategy shifts, attentional breakdown, and cognitive recalibration.

In addition to breakdown studies within a single system, CIS also calls for comparative methodologies that analyze the differential effects of distinct cognitive infrastructures across platforms, populations, and cultural contexts. This comparative strand is essential for understanding how design differences—such as opacity levels, algorithmic logic, personalization intensity, and data provenance—translate into distinct cognitive and epistemic consequences.

One such method involves cross-platform experimental comparisons, where matched participant groups interact with different AI-mediated under controlled task conditions. Researchers then measure variations in attention allocation (via eye-tracking), reasoning quality, decision-making accuracy, and information diversity. These studies aim to capture how different infrastructural logics promote or constrain cognitive diversity, epistemic access, and critical reflection.

Another approach employs naturalistic quasi-experiments by leveraging platform changes or policy updates as *naturally occurring breakdowns*. For instance, algorithmic changes that affect news feed composition or search result ranking can be tracked longitudinally using digital trace data, sentiment analysis, or user behavior metrics. Such analyses are especially powerful when paired with pre/post-intervention surveys, lab-based replications, or in-depth interviews to triangulate findings across levels of analysis.

These paradigms are designed not merely to detect changes in performance, but to measure the depth of cognitive coupling between human agents and infrastructural systems. In doing so, they help differentiate between superficial tool usage—where humans retain cognitive autonomy—and deep infrastructural integration, where AI systems have become functionally embedded within the user's cognitive architecture. Such methodologies enable the empirical study of System 0 effects, revealing how AI technologies operate as preconscious mediators of attention, reasoning, and judgment. They also offer a powerful lens for identifying cognitive dependencies, especially among different populations, digital literacies, or sociotechnical contexts.

Additionally, CIS demands multi-scale measurement capabilities that can track how individual cognitive changes aggregate into collective patterns. This requires integration of individual-level cognitive assessment, social network analysis of information flow, and cultural analysis of emerging meaning systems. Computational methods become essential for analyzing the vast data traces generated by cognitive infrastructures, but these must be complemented by qualitative approaches that can interpret the cultural significance of algorithmic transformations.

## 6. Conclusions

The pervasive influence of AI as cognitive infrastructure, or System 0, fundamentally reshapes human cognition, collective reasoning, and societal functioning in invisible yet foundational ways. This transformation is exemplified by individual cognitive dependencies, fragmented public spheres, and complex policy dilemmas, highlighting the urgent need for new frameworks to understand and govern these systems. Cognitive Infrastructure Studies (CIS) offers such a paradigm, reconceptualizing AI not merely as tools, but as embedded, constitutive layers of contemporary thought itself.

CIS aligns with traditional infrastructure studies by recognizing AI's embeddedness, modularity, and standardizing effects, but uniquely emphasizes the epistemic and normative consequences of AI-driven mediation. It extends the concept of the "extended mind" to "collective cognition," acknowledging that digital infrastructures now underpin societal decision-making, cultural transmission, and public discourse.

The implications are profound: just as physical infrastructures are regulated as public utilities, so too must cognitive infrastructures be interrogated and governed due to their role in structuring fundamental cognitive capacities, shaping information environments, generating behavioral dependencies, and amplifying network effects.

This necessitates a shift in digital governance from regulating individual AI applications to managing the underlying cognitive architectures that define how societies know, decide, and evolve. New analytic tools, empirical research, and policy frameworks are crucial to account for System 0's pervasive role in mediating cognition and power. Ultimately, CIS provides a conceptually ambitious yet socially grounded framework for understanding AI's societal impact, aiming to advance digital social research that illuminates how technologies mediate social practices, knowledge systems, and collective futures.

Future research must address critical questions about cognitive sovereignty: How can individuals and communities maintain autonomy over their thinking processes in algorithmically mediated environments? What design principles can ensure cognitive infrastructures enhance rather than constrain human intellectual capabilities? How can democratic societies govern the architectures of attention and relevance that increasingly shape public discourse?

The stakes could not be higher. As AI systems become more sophisticated at predicting and shaping human cognition, we face a choice between futures where cognitive infrastructures serve human flourishing or where they optimize for metrics disconnected from genuine human wellbeing. CIS provides both the conceptual framework and methodological tools necessary to ensure this choice is made consciously and democratically rather than by algorithmic default.

## 7. Funding Declaration

This research was partially supported by the Italian National Recovery and Resilience Plan (PNRR) under the Future Artificial Intelligence Research (FAIR) program, Project "Co-XAI: Cognitive Decision Intelligence Framework for Explainable AI Systems."

## 8. Data Availability

This article does not analyse or generate any dataset.

## 9. Ethical Statement

This article does not contain any studies with human participants performed by any of the authors.

## 10. References

Ananny M, & Crawford, K (2018) Seeing without knowing: Limitations of the transparency ideal and its application to algorithmic accountability. New Media & Society, *20*(3), 973-989. https://doi.org/10.1177/1461444816676645

Antikythera S (2025) Cognitive Infrastructures. Conjectural Explorations of AI as a Physical Actor in the Wild. Antikythera: Journal for the Philosophy of Planetary Computation. https://doi.org/10.1162/ANTI.5CZG

Baxter G, & Sommerville, I (2010) Socio-technical systems: From design methods to systems engineering. Interacting with Computers, *23*(1), 4-17. https://doi.org/10.1016/j.intcom.2010.07.003

Benkler Y (2006) The Wealth of Networks. How Social Production Transforms Markets and Freedom. Yale University Press. http://www.jstor.org/stable/j.ctt1njknw

Binz M, Akata, E, Bethge, M, Brändle, F, Callaway, F, Coda-Forno, J, Dayan, P, Demircan, C, Eckstein, MK, Éltető, N, Griffiths, TL, Haridi, S, Jagadish, AK, Ji-An, L, Kipnis, A, Kumar, S, Ludwig, T, Mathony, M, Mattar, M, Modirshanechi, A, Nath, SS, Peterson, JC, Rmus, M, Russek, EM, Saanum, T, Schubert, JA, Schulze Buschoff, LM, Singhi, N, Sui, X, Thalmann, M, Theis, FJ, Truong, V, Udandarao, V, Voudouris, K, Wilson, R, Witte, K, Wu, S, Wulff, DU, Xiong, H, & Schulz, E (2025) A foundation model to predict and capture human cognition. Nature. https://doi.org/10.1038/s41586-025-09215-4

Bommasani R, Arora, S, Chayes, J, Choi, Y, Cuéllar, M-F, Fei-Fei, L, Ho, DE, Jurafsky, D, Koyejo, S, Lakkaraju, H, Narayanan, A, Nelson, A, Pierson, E, Pineau, J, Singer, S, Varoquaux, G, Venkatasubramanian, S, Stoica, I, Liang, P, & Song, D (2025) Advancing science- and evidence-based AI policy. Science, *389*(6759), 459-461. https://doi.org/doi:10.1126/science.adu8449

Bowker GC, & Star, SL (1999) Sorting things out: Classification and its consequences. MIT Press.

Bucher T (2018) If... then: Algorithmic power and politics. Oxford University Press.

Callon M (1986) Some elements of a sociology of translation: Domestication of the scallops and the fishermen of St. Brieuc Bay. In Law J (Ed.), *Power, action and belief: A new sociology of knowledge?* (pp. 196–233). Routledge.

Chen M-Y, Lughofer, E, Rubio, JdJ, & Wu, YJ (2021) Editorial: Anticipatory Systems: Humans Meet Artificial Intelligence [Editorial]. Frontiers in psychology, *Volume 12 - 2021*. https://doi.org/10.3389/fpsyg.2021.721879

Chiriatti M, Ganapini, M, Panai, E, Ubiali, M, & Riva, G (2024) The case for human–AI interaction as system 0 thinking. Nature Human Behaviour, *8*(10), 1829-1830. https://doi.org/10.1038/s41562-024-01995-5

Chiriatti M, Ganapini, M, Panai, E, Wiederhold, BK, & Riva, G (2025) System 0: Transforming Artificial Intelligence into a Cognitive Extension. Cyberpsychology, Behavior & Social Networking, *28*(7), 534-542. https://doi.org/10.1089/cyber.2025.0201

Clark A, & Chalmers, D (1998) The extended mind. Analysis, *58*(1), 7-19.

Coeckelbergh M (2025) AI and Epistemic Agency: How AI Influences Belief Revision and Its Normative Implications. Social Epistemology, 1-13. https://doi.org/10.1080/02691728.2025.2466164

Cole M, & Engestrom, J (1993) A cultural-historical approach to distributed cognition. In Salomon G (Ed.), *Distributed cognitions* (pp. 1-46). Cambridge University Press.

Couldry N, & Hepp, A (2018) The mediated construction of reality. John Wiley & Sons.

Edwards PN, Jackson, SJ, Bowker, GC, & Knobel, CP (2007) Understanding Infrastructure: Dynamics, Tensions, and Design History and Theory of Infrastructure: Lessons for New Scientific Cyberinfrastructures, Ann Arbor.

Flaxman S, Goel, S, & Rao, JM (2016) Filter Bubbles, Echo Chambers, and Online News Consumption. Public Opinion Quarterly, *80*(S1), 298-320. https://doi.org/10.1093/poq/nfw006

Floridi L (2025) AI as Agency without Intelligence: On Artificial Intelligence as a New Form of Artificial Agency and the Multiple Realisability of Agency Thesis. Philosophy & Technology, *38*(1), 30. https://doi.org/10.1007/s13347-025-00858-9

Fraser N (1990) Rethinking the Public Sphere: A Contribution to the Critique of Actually Existing Democracy. Social Text(25/26), 56-80. https://doi.org/10.2307/466240

Glickman M, & Sharot, T (2025) How human–AI feedback loops alter human perceptual, emotional and social judgements. Nature Human Behaviour(9), 345-359. https://doi.org/10.1038/s41562-024-02077-2

Habermas J (1991/1962) The structural transformation of the public sphere: An inquiry into a category of bourgeois society. MIT Press.

Hollan J, Hutchins, E, & Kirsh, D (2000) Distributed cognition: toward a new foundation for human-computer interaction research. ACM Trans. Comput.-Hum. Interact., *7*(2), 174–196. https://doi.org/10.1145/353485.353487

Hutchins E (1995) Cognition in the wild. MIT Press.

Kahneman D (2011) Thinking, fast and slow. Macmillan.

Kramer AD, Guillory, JE, & Hancock, JT (2014) Experimental evidence of massive-scale emotional contagion through social networks. Proc Natl Acad Sci U S A, *111*(24), 8788-8790. https://doi.org/10.1073/pnas.1320040111

Latour B (2005) Reassembling the social: An introduction to actor-network-theory. Oxford University Press.

Law J, & Hassard, J (Eds.). (1999) *Actor network theory and after.* Blackwell.

Li Vigni F (2021) Complexity Sciences: A Scientific Platform. Science & Technology Studies, *34*(4), 30-55. https://doi.org/10.23987/sts.97027

Lorenz-Spreen P, Oswald, L, Lewandowsky, S, & Hertwig, R (2023) A systematic review of worldwide causal and correlational evidence on digital media and democracy. Nature Human Behaviour, *7*(1), 74-101. https://doi.org/10.1038/s41562-022-01460-1

Mercier H, & Sperber, D (2011) Why do humans reason? Arguments for an argumentative theory. Behavioral and Brain Sciences, *34*(2), 57-74.

Mercier H, & Sperber, D (2017) The enigma of reason (Vol. 12). Harvard University Press. https://doi.org/https://doi.org/10.14763/2023.1.1683

Milan S (2015) When algorithms shape collective action: Social media and the dynamics of cloud protesting. Social Media+ Society, *1*(2), 2056305115622481.

Mittelstadt BD, Allo, P, Taddeo, M, Wachter, S, & Floridi, L (2016) The ethics of algorithms: Mapping the debate. Big Data & Society, *3*(2), 2053951716679679. https://doi.org/10.1177/2053951716679679

Morgan TJH, & Feldman, MW (2024) Human culture is uniquely open-ended rather than uniquely cumulative. Nature Human Behaviour. https://doi.org/10.1038/s41562-024-02035-y

Nguyen CT (2020) Echo Chambers and Epistemic Bubbles. Episteme, *17*(2), 141-161. https://doi.org/10.1017/epi.2018.32

Nieminen JH, & Ketonen, L (2024) Epistemic agency: a link between assessment, knowledge and society. Higher Education, *88*(2), 777-794. https://doi.org/10.1007/s10734-023-01142-5

NMI (2025) Emotional risks of AI companions demand attention. Nature Machine Intelligence, *7*(7), 981-982. https://doi.org/10.1038/s42256-025-01093-9

Plantin J-C, Lagoze, C, Edwards, PN, & Sandvig, C (2018) Infrastructure studies meet platform studies in the age of Google and Facebook. New Media & Society, *20*(1), 293-310. https://doi.org/10.1177/1461444816661553

Reckwitz A (2002) Toward a Theory of Social Practices:A Development in Culturalist Theorizing. European Journal of Social Theory, *5*(2), 243-263. https://doi.org/10.1177/13684310222225432

Risko EF, & Gilbert, SJ (2016) Cognitive Offloading. Trends in Cognitive Sciences, *20*(9), 676-688. https://doi.org/10.1016/j.tics.2016.07.002

Riva G (2025a) Digital 'We': Human Sociality and Culture in the Era of Social Media and AI. American Psychologist. https://doi.org/10.1037/amp0001577

Riva G, Wiederhold, BK, & Cipresso, P (2025) Humane Artificial Intelligence: Psychological, Social, and Ethical Dimensions. Cyberpsychology, Behavior, and Social Networking, *28*(7), 457-461. https://doi.org/10.1089/cyber.2025.0202

Salvi F, Horta Ribeiro, M, Gallotti, R, & West, R (2025) On the conversational persuasiveness of GPT-4. Nature Human Behaviour, *9*(8), 1645-1653. https://doi.org/10.1038/s41562-025-02194-6

Schatzki TR (2002) The site of the social: A philosophical account of the constitution of social life and change. Pennsylvania State University Press.

Star SL (1999) The Ethnography of Infrastructure. American Behavioral Scientist, *43*(3), 377-391. https://doi.org/10.1177/00027649921955326

Star SL, & Ruhleder, K (1996) Steps Toward an Ecology of Infrastructure: Design and Access for Large Information Spaces. Information Systems Research, *7*(1), 111-134. https://doi.org/10.1287/isre.7.1.111

Susser D (2019) Invisible Influence: Artificial Intelligence and the Ethics of Adaptive Choice Architectures Proceedings of the 2019 AAAI/ACM Conference on AI, Ethics, and Society, Honolulu, HI, USA. https://doi.org/10.1145/3306618.3314286

van Dijck J, Poell, T, & de Waal, M (2018) The Platform Society. Oxford University Press. https://doi.org/10.1093/oso/9780190889760.001.0001

Verbeek P-P (2011) Moralizing technology: Understanding and designing the morality of things. University of Chicago Press.

Werner T, Soraperra, I, Calvano, E, Parkes, DC, & Rahwan, I (2024) Experimental evidence that conversational artificial intelligence can steer consumer behavior without detection. arXiv preprint arXiv:2409.12143.